\documentclass[aps,pop,twocolumn,showpacs,superscriptaddress,article]{revtex4}
\usepackage{mathrsfs}
\usepackage{amsmath}
\usepackage{bm}
\usepackage{color}
\usepackage{graphicx}
\usepackage{subfig}
\usepackage{hyperref}
\begin{document}
%\linenumbers
\title{Excitation of Zonal Flow by Intermediate-Scale Toroidal  Electron Temperature Gradient Turbulence}

\author{Haotian Chen\footnote{Permanent address: Institute of Space Science and Technology, Nanchang University, Nanchang, 330031, People's Republic of China}}\email[]{Email: haotianchen-ext@us.es}
\affiliation{University of Colorado at Boulder, Boulder, Colorado 80309, USA}
\author{Stefan Tirkas}
\affiliation{University of Colorado at Boulder, Boulder, Colorado 80309, USA}
\author{Scott E. Parker}
\affiliation{University of Colorado at Boulder, Boulder, Colorado 80309, USA}

%\vskip 0.25cm

\date{\today}

\begin{abstract}
	We show that zonal flow can be preferentially excited by intermediate-scale toroidal electron temperature gradient (ETG) turbulence in tokamak plasmas.
	Previous theoretical studies that yielded an opposite conclusion assumed a fluid approximation for ETG modes.
	Here, we carry out a gyrokinetic analysis which ultimately yields a nonlinear Schr\"{o}dinger equation for the ETG dynamics with a Navier-Stokes type nonlinearity.
	%The threshold condition of the modulational instability and associated saturation level are estimated. 
	For typical tokamak parameters, it is found that zonal flow generation plays an important role in the intermediate-scale ETG turbulence.
	This finding offers an explanation for recent multi-scale gyrokinetic simulations.

\end{abstract}


\maketitle

%\section{Introduction}
%\label{sec:may:01:16:01}
%\emph{Introduction.}---
%An outstanding issue in magnetically confined plasmas is the mechanism of anomalous electron energy transport. It is of particular importance for future burning plasmas such as ITER because fast alpha particles mostly heat electrons.
%The electron temperature gradient (ETG) driven turbulence has been proposed as a promising candidate, since there is little zonal flow generation and  the radially extended linear streamers can persist in the nonlinear saturated state \cite{kim03, lin05, chen05, nevins}.
%\begin{linenumbers}
Recently, large-box-size, long-time-scale gyrokinetic simulations have been  carried out to investigate the toroidal electron temperature gradient (ETG) driven turbulence \cite{parker,howard16,colyer, holland17}.
It is found that, unlike the usual theoretical expectations \cite{kim03, lin05, chen05}, the formation of  zonal flow (ZF) can play an important role in regulating intermediate-scale ETG turbulence and become dominant in the final saturation. 
Since the long-time saturated state is more experimentally relevant, ETG-ZF dynamics is crucial for understanding the nonlinear ETG physics and associated turbulent transport.
In this work, motivated by the simulation observations, we present a gyrokinetic analysis addressing the spontaneous and forced ZF generation on the same footing.
Our results indicate that the ETG nonlinearity is of a Navier-Stokes form at intermediate-scales, which is generally much stronger than the Hasegawa-Mima type nonlinearity in fluid limit \cite{lin05, chen05}, and, thus, a significant ZF generation is expected for intermediate-scale toroidal ETG turbulence.

For simplicity and clarity, we consider an axisymmetric, low-$\beta$ (plasma to magnetic pressure), large aspect-ratio ($\epsilon=r/R_{0}\ll 1$) tokamak with the usual  minor radius ($r$), poloidal ($\theta$) and toroidal ($\zeta$) coordinates.
%Thus we can employ here the $s-\alpha$ model \cite{connor78}  with $\alpha=0$ for equilibrium the magnetic field.
%In the weak turbulence regime, the electrostatic fluctuation is taken to be coherent and consists of a high-$n$ ETG mode and a zonal perturbation.
We examine a single  high-$n$ ETG mode and associated zonal flow.
%Furthermore, to explicitly account for the separation in spatio-temporal scales, 
Adopting ballooning representation \cite{connor78}, the ETG fluctuation can be written as
\begin{eqnarray}
\label{eq:pump}
	\delta\phi_{\bm{k}}=\sum_{m}e^{i(m\theta-n\zeta)}\iint d\eta d\theta_{k}e^{i[nq(\eta-\theta_{k})-m\eta]}A_{\bm{k}}\delta\tilde{\phi}_{\bm{k}},\nonumber
\end{eqnarray}
%and
%\begin{eqnarray}
%\label{eq:sidebands}
	%\delta\phi_{\pm}=A_{\pm}\sum_{m}e^{i(m\theta-n\zeta \mp\int n\theta_{z}dq-\omega_{\pm}t)}\int e^{i(nq-m)\eta}d\eta \delta \tilde{\phi}_{\pm},\nonumber
%\end{eqnarray}
where the subscript $\bm{k}\equiv(n,\theta_{k})$ denotes the wavenumber space, and $nq'\theta_{k}$ is a radial envelope wavenumber with $q(r)$ being the safety factor. 

The analysis presented here will focus on intermediate-scale turbulence with $k_{\perp}^{2}\rho_{e}^{2}\ll 1\ll k_{\perp}^{2}\rho_{i}^{2}$, where the Debye shielding is negligible, and the ETG mode is nearly isomorphic to its ion-scale counterpart ITG mode, except for the ion adiabatic response for both ZF and ETG.
We can therefore impose the quasineutrality condition
\begin{eqnarray}
\label{eq:quasineutrality}
	(1+\tau)\Phi_{\bm{k}}+\langle J_{\bm{k}}\delta H_{\bm{k}}\rangle_{v}=0,
\end{eqnarray}
where $\tau=T_{e}/T_{i}$ is the temperature ratio, $J_{\bm{k}}=J_{0}(k_{\perp}\rho_{e}v_{\perp})$ is the zeroth-order Bessel function accounting for the finite Larmor radius (FLR) effect, and $\delta H_{\bm{k}}$  can be derived from the nonlinear gyrokinetic equation \cite{frieman}:
\begin{eqnarray}
\label{eq:nlgk}
	& &\mathcal{L}_{\bm{k}}\delta H_{\bm{k}}+(i\partial_{t}+\omega_{*e}^{t})F_{0}J_{\bm{k}}\Phi_{\bm{k}}\nonumber\\
	&=&\frac{i\rho^{2}_{e}v_{te}}{2r_{n}}\sum_{\bm{k}_{2}}\{[J_{\bm{k}_{1}}\Phi_{\bm{k}_{1}},\delta H_{\bm{k}_{2}}^{*}]+[J_{\bm{k}_{2}}\Phi_{\bm{k}_{2}}^{*},\delta H_{\bm{k}_{1}}]\}.
\end{eqnarray}
Here, 
we have normalized the electrostatic potential as $\Phi_{\bm{k}}=e\delta\phi_{\bm{k}}/(\rho_{*}T_{e})$ with $\rho_{*}=\rho_{e}/r_{n}$, and defined the Poisson bracket $[A, B]=(\partial_{\theta}A/r)\partial_{r}B-(\partial_{r}A)\partial_{\theta}B/r$.
 $\mathcal{L}_{\bm{k}}=[i\partial_{t}+\omega_{t}+\omega_{d}]$ is the inverse phase-space linear  propagator, where $\omega_{t}=v_{te}v_{\parallel}(n q+i\partial_{\theta})/(q R_{0})$ and $\omega_{d}=(v_{\parallel}^{2}+v_{\perp}^{2}/2)v_{te}^{2}(i\sin\theta \partial_{r}-k_{\theta}\cos\theta)/(|\omega_{ce}|R_{0})$ denote, respectively, the transit and magnetic drift frequencies.
$\omega_{*e}=k_{\theta}c T_{e}/(eB r_{n})$ is the electron diamagnetic drift frequency, $\omega_{*e}^{t}=-\omega_{*e}[1+\eta_{e}(v^{2}-3/2)]$ and $\eta_{e}=r_{n}/r_{ti}$, with $k_{\theta}=m/r$, and $r_{n}$ and $r_{t_{i}}$ being, respectively, the equilibrium density and temperature scale lengths. 
$F_{0}$ is the local Maxwellian. $\bm{k}_{2}$ satisfies the matching conditions $\bm{k}=\bm{k}_{1}-\bm{k}_{2}$. 
%Other notations are customary. 
%Equations (\ref{eq:quasineutrality}) and (\ref{eq:nlgk}) form the closed set of equations for the ETG-ZF coherent nonlinear system.

\begin{figure}[!htp]
\vspace{-0.3cm}
\setlength{\belowcaptionskip}{-0.1cm}
\centering
\includegraphics[scale=0.35]{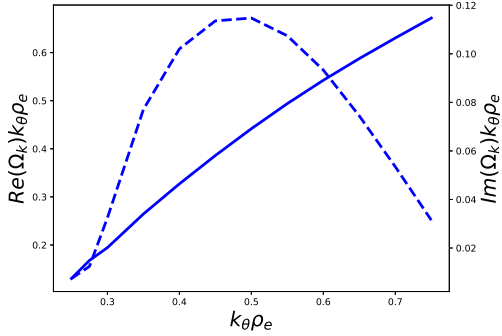}
	\caption{(Color online) Normalized growth rate (dashed line) and real frequency (solid line) vs $k_{\theta}\rho_{e}$ with $\theta_{k}=0$, $\tau=1$, $\epsilon=0.25$, $\epsilon_{n}=0.3$, $\overline{\eta}=1.5$, $\eta_{e}=3$, $q=1.4$ and $s=0.8$.}
\label{eps:dispersion}
\end{figure}

%\emph{Linear properties.}---
In ballooning space, it is well known that the dominant order linear gyrokinetic equation is an ordinary differential equation parameterized by $\theta_{k}$.
To allow a tractable weak turbulence analysis, we assume a local kinetic model \cite{kim} to capture essential linear properties of toroidal ETG mode.
In particular, we replace the $\eta$ variable by its parallel-mode-structure-averaged value $\bar{\eta}=[\int d\eta \tilde{\Phi}^{*}\eta^{2}\tilde{\Phi}]^{1/2}$.
%Such an approximation is justified for the toroidal ETG mode with a moderate ballooning structure.
The linear dispersion relation then reduces to an algebraic equation
\begin{eqnarray}
\label{eq:dispersion}
	D_{\bm{k}}(\omega_{\bm{k}},\theta_{k},r)\equiv(1+\tau)-\langle \frac{(\omega_{\bm{k}}+\omega^{t}_{*e})\bar{J}^{2}_{\bm{k}}F_{0}}{\omega_{\bm{k}}+\omega_{t}+\omega_{d}}\rangle_{v}=0,
\end{eqnarray}
where $\omega_{t}=-v_{te}v_{\parallel}/(q R_{0} \bar{\eta})$ and $\omega_{d}=-\omega_{*e}\epsilon_{n}(2v^{2}_{\parallel}+v^{2}_{\perp}) [s(\bar{\eta}-\theta_{k})\sin\bar{\eta}+\cos\bar{\eta}]$, with the magnetic shear $s=rq'/q$ and $\epsilon_{n}=r_{n}/R_{0}$. 
$\langle\cdots\rangle_{v}\equiv\int d\bm{v}(\cdots)$ and the $r$ dependency corresponds to equilibrium variations,
As an example, the normalized linear ETG spectrum $\Omega_{\bm{k}}=\omega_{\bm{k}}/|\omega_{*e}|$ for typical tokamak plasma parameters is plotted in Fig. (\ref{eps:dispersion}).
%\end{linenumbers}
%Note that the linear eigenmode equation can be straightforwardly extended to a general ballooning model\cite{Romanelli}, however, such theory is more cumbersome and beyond the scope of the present study.

%\emph{Weak Turbulence Theory.}---
%Since ions can be treated as adiabatic,  only electrons contribute to the nonlinear physics. 
%\begin{linenumbers}
The nonlinear description of ZF can be obtained by taking neoclassical effects into account \cite{rosenbluth,kim03}, yielding
\begin{eqnarray}
\label{eq:nonlocal_Az}
	& &[\partial_{t}+\gamma_{z}(1+d_{z}k_{\theta}^{2}\rho_{e}^{2}s^{2}\theta_{k}^{2})]\chi_{z} A_{z}(\theta_{k})\nonumber\\
	&=&\sqrt{\frac{\pi}{2}} (k_{\theta}\rho_{e}s)^{3}\theta_{k}\int d\vartheta_{k} \vartheta^{2}_{k} [A_{n}(\vartheta_{k})A_{n}^{*}(\vartheta_{k}-\theta_{k})a_{n}^{*}\nonumber\\
	& &-A_{n}(\vartheta_{k}+\theta_{k})A^{*}_{n}(\vartheta_{k})a_{n}],
\end{eqnarray}
where  the length and time scales are normalized to $\rho_{e}$ and $|\omega_{*e}^{-1}|$, respectively. 
We have added an \emph{ad hoc} gyrodiffusive contribution ($\propto d_{z}$) to the ZF collisional damping rate, of which the importance has been emphasized previously \cite{ricci}. 
$\gamma_{z}\simeq 3\nu_{ee}/(|\omega_{*e}|\sqrt{\epsilon})$ with $\nu_{ee}$ being the electron-electron collision frequency, and $\chi_{z}=\tau+(1+1.6q^{2}/\sqrt{\epsilon})k_{\theta}^{2}\rho_{e}^{2}s^{2}\theta_{k}^{2}/2$ is the total susceptibility  \cite{kim03}. 
%in which the $\propto(1+1.6q^{2}/\sqrt{\epsilon})$ term comes from the neoclassical polarization.
The nonlinear term arises from Reynolds stress, where we defined a parallel decoupling  function $a_{n}=a_{n}(\theta_{k},\vartheta_{k})=\int d\eta \langle \tilde{\Phi}^{*}_{n}(\eta,\vartheta_{k})v_{\perp}^{2}\delta \tilde{H}_{n}(\eta,\vartheta_{k}+\theta_{k})\rangle_{v}$  to measure the parallel correlation of ETG turbulence, and it can limit the nonlinearity via the finite localization of linear parallel mode structures.
As a simple but relevant paradigm, we assume $a_{n}=a_{n}(0,0)\textrm{exp}(-\theta_{k}^{2}/2\bar{\eta}^{2})$ in the present study, which essentially treats $\theta_{k}$ as a tilting angle \cite{sugama,chen18}.

Solving Eq. (\ref{eq:nlgk}) to the next order, the quasineutrality condition straightforwardly produces the following nonlinear Schr\"{o}dinger equation for the ETG amplitude, 
\begin{eqnarray}
\label{eq:nonlocal_An}
	& &[i(\partial_{t}-\gamma_{n})-b_{n}k_{\theta}^{2}\rho_{e}^{2}s^{2}\theta_{k}^{2}-\frac{c_{n}}{k_{\theta}^{2}\rho_{e}^{2}s^{2}}\frac{\partial^{2}}{\partial \theta_{k}^{2}}]A_{n}(\theta_{k})\nonumber\\
	&=&-\frac{ik_{\theta}\rho_{e}s}{\sqrt{2\pi}}\int d\vartheta_{k} \vartheta_{k} A_{z}(\vartheta_{k} )A_{n}(\theta_{k} -\vartheta_{k}).
\end{eqnarray}
Here, $\gamma_{n}$ is the linear growth rate, $b_{n}k_{\theta}^{2}\rho_{e}^{2}s^{2}\theta_{k}^{2}$ denotes the frequency mismatch, and the $\propto c_{n}$ term recovers the correction associated with the plasma nonuniformities in real space. 
The parameters $b_{n}=-(\partial_{\theta_{k}}^{2}D_{\bm{k}}/\partial_{\Omega_{\bm{k}}}D_{\bm{k}})/(2k_{\theta}^{2}\rho_{e}^{2}s^{2})$ and $c_{n}=(\partial_{r}^{2}D_{\bm{k}}/\partial_{\Omega_{\bm{k}}}D_{\bm{k}})/2$ are determined by linear ETG dynamics.
The nonlinear term, meanwhile,  is the coupling of ZF to ETG mode via $E_{r}$ shearing. 
One readily identifies that, the ETG saturation is set by competition between linear growth and ZF-induced scattering to the linearly stable short radial wavelengths.

Equations (\ref{eq:nonlocal_Az}) and (\ref{eq:nonlocal_An}), along with the complex parameters solely by linear ETG properties, fully characterize the dynamics of coherent ETG-ZF system, and will hereafter be referred to as the nonlinear Schr\"{o}dinger equation (NLSE) model.
To properly account for the kinetic effects, the conventional fluid limit is not assumed here, while both the forced and spontaneous generation of ZF are kept on the equal footing. 
%The importance of forced generation of ZF has also been observed in an earlier simulation work \cite{parker}.
%Kinetic effects are known to play significant roles in the linear features of long-wavelength ETG modes.
We emphasize that the coupling of ZF to ETG is formally of Hasegawa-Mima type in the fluid limit \cite{hasegawa, chen05}, and is $\mathcal{O}(k_{\theta}^{2}\rho_{e}^{2})$ weaker than the Navier-Stokes type nonlinearity in the present gyrokinetic analysis. As a consequence, ZF can more easily regulate the underlying ETG turbulence, and it will be shown later that the threshold condition for spontaneous ZF excitation  is reduced by a factor $\mathcal{O}(k_{\theta}^{2}\rho_{e}^{2})$ relative to previous fluid prediction. 
%This suggests that the ZF may have special importance in the intermediate-scale ETG turbulence.

%Before presenting detailed analyses of Eqs. (\ref{eq:nonlocal_Az}) and (\ref{eq:nonlocal_An}), 
It is illuminating to notice that a four-wave model \cite{chen00} can be straightforwardly extracted from the NLSE model.
By ignoring plasma nonuniformities and assuming the narrow-band ZF and ETG amplitudes, respectively, 
as $A_{z}\Pi[(\theta_{k}-\theta_{z})/W]$ and $A_{0}\Pi(\theta_{k}/W)+A_{+}\Pi[(\theta_{k}-\theta_{z})/W]+A_{-}\Pi[(\theta_{k}+\theta_{z})/W]$, one readily obtains
\begin{eqnarray}
\label{eq:local_Az}
	& &[\partial_{t}+\gamma_{z}(1+d_{z}k_{z}^{2}\rho_{e}^{2})]\chi_{z}A_{z}\nonumber\\
	&=&\sqrt{\pi/2} W k_{z}^{3}\rho_{e}^{3}(A_{+}A^{*}_{0}a_{n}^{*}-A^{*}_{-}A_{0}a_{n}),
\end{eqnarray}
\begin{eqnarray}
\label{eq:local_Ap}
	[\partial_{t}+ i\Delta-\gamma_{s}]A_{+}=-W k_{z}\rho_{e}A_{z}A_{0}/\sqrt{2\pi},
\end{eqnarray}
\begin{eqnarray}
\label{eq:local_Am}
	[\partial_{t}+ i\Delta-\gamma_{s}]A_{-}=W k_{z}\rho_{e}A^{*}_{z}A_{0}/\sqrt{2\pi},
\end{eqnarray}
and 
\begin{eqnarray}
\label{eq:local_A0}
	[\partial_{t}-\gamma_{n}]A_{0}=-W k_{z}\rho_{e}( A_{z}A_{-}-A^{*}_{z}A_{+})/\sqrt{2\pi}.
\end{eqnarray}
Here, $\Pi$ is the usual rectangle function, $W$ explicitly denotes the bandwidth, $k_{z}=k_{\theta}s\theta_{z}$ is the radial wavenumber, and $A_{0}$ and $A_{\pm}$ are, respectively, the pump ETG mode and sidebands produced by the envelope modulation. $\Delta=\textrm{Re}(b_{n})k_{z}^{2}\rho_{e}^{2}$ is the frequency mismatch and $\gamma_{s}=\gamma_{n}+\textrm{Im}(b_{n})k^{2}_{z}\rho_{e}^{2}$ is the linear growth/damping rate of sidebands.
The four-wave model has the following the conservation property
\begin{eqnarray}
\label{eq:conservation}
	(\partial_{t}-2\gamma_{n})|A_{0}|^{2}=(2\gamma_{s}-\partial_{t})(|A_{+}|^{2}+|A_{-}|^{2}).
\end{eqnarray}

\begin{figure}[!htp]
\vspace{-0.3cm}
\setlength{\belowcaptionskip}{-0.1cm}
\centering
\includegraphics[scale=0.4]{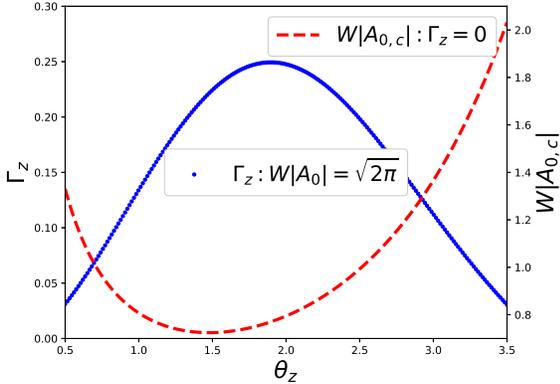}
	\caption{(Color online) Normalized critical threshold amplitude $W|A_{0,c}|$ and ZF growth rate $\Gamma_{z}$ versus $\theta_{z}$ for $k_{\theta}\rho_{e}=0.3$,  $\gamma_{z}=0.025$ and $d_{z}=2$. Equation (\ref{eq:dispersion})  yields $a_{n}(0,0)\simeq-3-2i$ and $b_{n}\simeq-2-2.5i$. The rest of the parameters are the same as Fig. (\ref{eps:dispersion}).}
\label{eps:A0C}
\end{figure}

The four-wave model is a dynamical system that displays both weak and strong nonlinear behaviours.
We first explore the onset condition of the modulational instability with a constant pump amplitude $A_{0}$.
In this case, the system is linear and a dispersion relation can be derived from Eqs. (\ref{eq:local_Az})-(\ref{eq:local_Am}), by letting $\partial_{t}\equiv\Gamma_{z}$, as
\begin{eqnarray}
\label{eq:nonlinear_dispersion}
	& &[(\Gamma_{z}-\gamma_{s})^{2}+\Delta^{2}][\Gamma_{z}+\gamma_{z}(1+d_{z}k_{z}^{2}\rho_{e}^{2})]\chi_{z}\nonumber\\
	&=&k^{4}_{z}\rho^{4}_{e}W^{2}|A_{0}|^{2}[\Delta\textrm{Im}(a_{n})-(\Gamma_{z}-\gamma_{s})\textrm{Re}(a_{n})],
\end{eqnarray}
which gives the critical threshold condition:
\begin{eqnarray}
\label{eq:A0c}
	W^{2}|A_{0,c}|^{2}=\frac{(\Delta^{2}+\gamma_{s}^{2})\gamma_{z}(1+d_{z}k_{z}^{2}\rho_{e}^{2})\chi_{z}}{k_{z}^{4}\rho_{e}^{4}[\gamma_{s}\textrm{Re}(a_{n})+\Delta\textrm{Im}(a_{n})]}.
\end{eqnarray}
Thus, as discussed earlier, the threshold pump wave intensity is much lower (an order $\mathcal{O}(k_{\theta}^{2}\rho_{e}^{2})$) than the value from fluid theory \cite{chen05}.
Moreover, after some straightforward algebra, one can show that the threshold $|A_{0,c}|^{2}$ is minimized at $\theta_{z,m}^{2}\simeq 4\bar{\eta}^{2}/\{1+[1+4\kappa^{2} d_{z}+4\kappa^{4}(1+1.6q^{2}/\sqrt{\epsilon})(d_{z}+1/\kappa^{2})/\tau ]^{1/2}\}$, with $\kappa^{2}=2k_{\theta}^{2}\rho_{e}^{2}s^{2}\bar{\eta}^{2}$.
For typical tokamak parameters,  as shown in Fig. (\ref{eps:A0C}), ZFs are more easily excited around $|\theta_{z,m}|\sim\mathcal{O}(1)$,  and the growth rate above the critical amplitude threshold,  meanwhile, peaks at $|\theta_{z}|\gtrsim |\theta_{z,m}|$.
The ETG-ZF interactions, thereby, tend to ultimately isotropize the linear streamers.
These effects are mainly attributed to the parallel decoupling between the pump and sidebands in Reynolds stress term, rather than the gyrodiffusive correction.
%Another distinguishing property of the threshold condition is the linear proportionality of $|A_{0,c}|^{2}$ to the zonal flow damping rate.
%\end{linenumbers}

%\begin{linenumbers}
Next, consider the temporal evolution of the four-wave model.
From Eq. (\ref{eq:conservation}), one can easily show that explosive growth exists for $\gamma_{n},\gamma_{s}>0$.
%Equation (\ref{eq:conservation}) demonstrates the existence of an explosive state for $0<\gamma_{n},\gamma_{s}$.
On the other hand, for $\gamma_{n}+\gamma_{s}<0$, we can demonstrate, by using Eqs. (\ref{eq:local_Az})-(\ref{eq:local_A0}), that the four-wave model gives a fixed-point solution with a constant $A_{z}$ for sufficiently low ZF damping rate, namely,
\begin{eqnarray}
\label{eq:fp_Az}
	k_{z}^{2}\rho_{e}^{2}W^{2}|A_{z,p}|^{2}=\pi(\delta^{2}-\Delta \delta-\gamma_{n}\gamma_{s}),
\end{eqnarray}
and
\begin{eqnarray}
\label{eq:fp_A0}
	k_{z}^{2}\rho_{e}^{2}W^{2}|A_{0,p}|^{2}=\frac{\chi_{z}\gamma_{z}(1+d_{z}k_{z}^{2}\rho_{e}^{2})[\gamma_{s}^{2}+(\delta-\Delta)^{2}]}{k_{z}^{2}\rho_{e}^{2}[(\Delta-\delta)\textrm{Im}(a_{n})+\gamma_{s}\textrm{Re}(a_{n})]},
\end{eqnarray}
where $\delta=\Delta\gamma_{n}/(\gamma_{s}+\gamma_{n})$ is the amplitude oscillation frequency of ETGs due to their nonlinear interplay with the ZF. That is, the ETG turbulence is still fluctuating as the ZF converges to a steady-state, consistent with recent numerical results \cite{colyer}.
It also follows from Eqs. (\ref{eq:fp_Az}) and (\ref{eq:fp_A0}) that, as in the case of ITG turbulence \cite{diamond98}, $|A_{0,p}|^{2}$ is proportional to the ZF collisional damping, while the ZF level is $\gamma_{z}$ independent.
However, it is important to note that the estimate of the saturated ETG fluctuation level given by Eq. (\ref{eq:fp_A0}) is only valid for the four-wave model with a single ZF mode.
For any realistic system with a spectrum of radial ZF modes, the ETG turbulence will subsequently continue driving ZF with lower threshold condition. 
%until the ETG fluctuation level asymptotically approaches the marginal stability characterized by the minimum $|A_{0,c}|^{2}$ value.
Thus, one may use the $|A_{0,p}|^{2}$ value at $\theta_{z}=\theta_{z,m}$  to quantitatively estimate the ETG saturation level.

%Therefore, for sufficiently small $\gamma_{z}$, we can anticipate that the four-wave model exhibits three kinds of nonlinear behaviors depending on the radial wavenumber $k_{z}^{2}\rho_{e}^{2}$: fixed points with $\gamma_{n}+\gamma_{s}<0$,  periodic oscillations with $\gamma_{n}+\gamma_{s}>0$ and $\gamma_{s}<0$, and explosive states with $\gamma_{n},\gamma_{s}>0$.
%However, similar to the ITG-ZF dynamics \cite{chen00}, as $\gamma_{z}$ increases to the high collisionality regime, the fixed point solution is destabilized, and the four-wave system first oscillates in limit cycle states, and then enters the chaotic regime. 
%For simplicity, these features are not present here.
%These features are illustrated in Fig. (\ref{eps:bif}), where the local extremas of $A_{z}$ are depicted after the transient phase. The measured fixed point solution $A_{z}\simeq 1.92$ agrees well with the analytical prediction from Eq. (\ref{eq:fp_Az}).

%\begin{figure}[!htp]
%\centering
%\includegraphics[scale=0.45]{bif3.eps}
	%\caption{(Color online) Bifurcation diagram with varying $\gamma_{z}$ for $k_{z}\rho_{e}=0.3$. The rest of parameters are the same as Fig. (\ref{eps:A0C}).}
%\label{eps:bif}
%\end{figure}

\begin{figure}[!htp]
\vspace{-0.3cm}
\setlength{\belowcaptionskip}{-0.1cm}
\centering
	\includegraphics[scale=0.35]{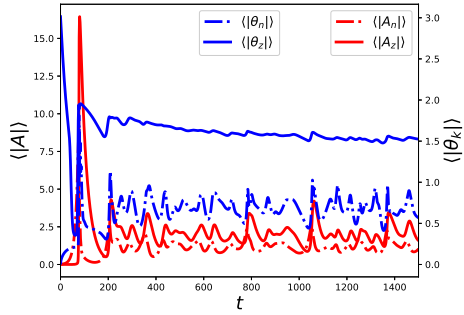}
	\caption{(Color online) Time histories of the averaged amplitudes and radial wavenumbers, for $\gamma_{z}=0.025$ and $c_{n}=(-1.25+7.5i)\times 10^{-5}$. The other parameters are the same as Fig. (\ref{eps:A0C}).}
\label{eps:history}
\end{figure}

\begin{figure}[!htp]
\vspace{-0.3cm}
\setlength{\belowcaptionskip}{-0.1cm}
\centering
	\includegraphics[scale=0.35]{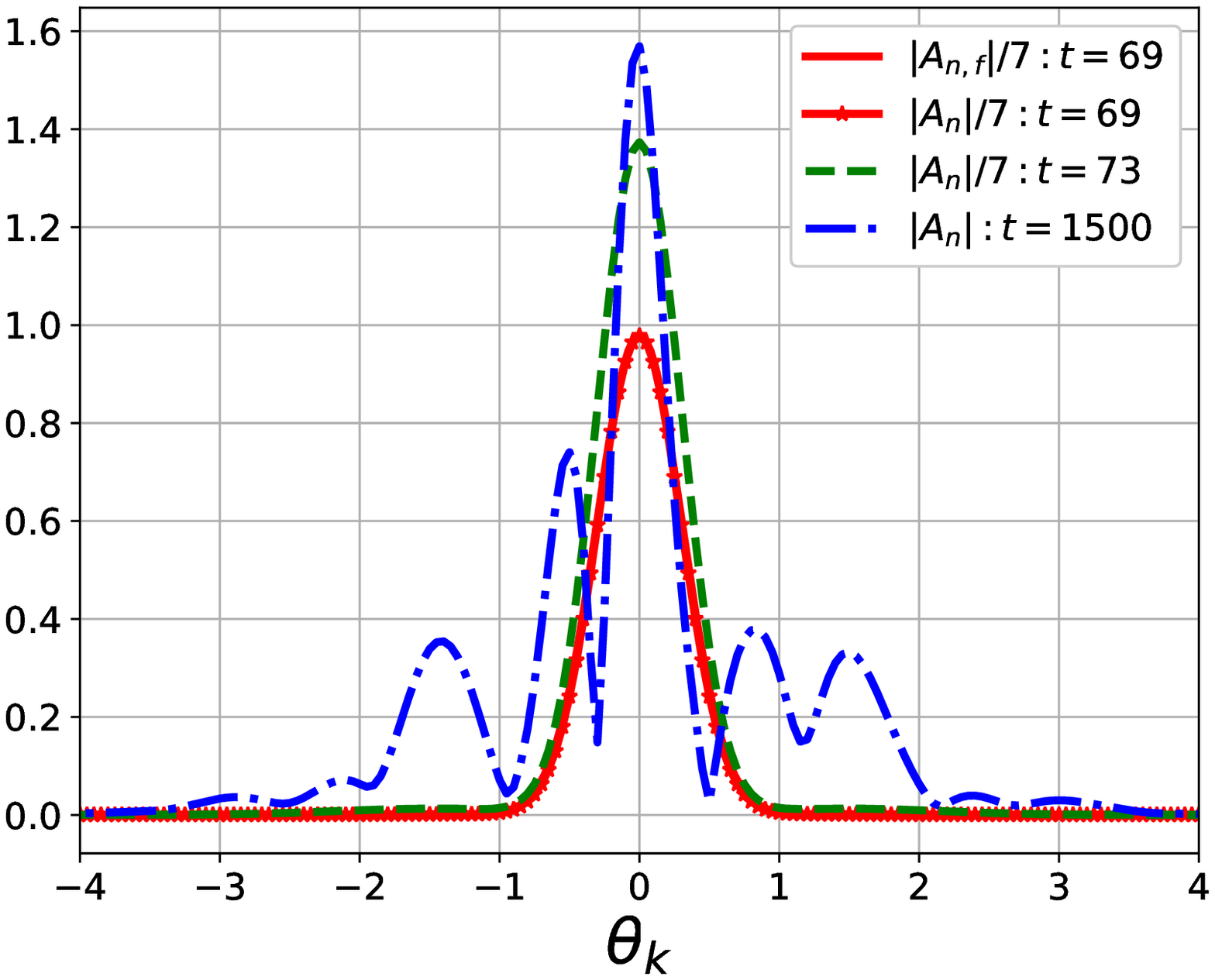}\\
	\includegraphics[scale=0.35]{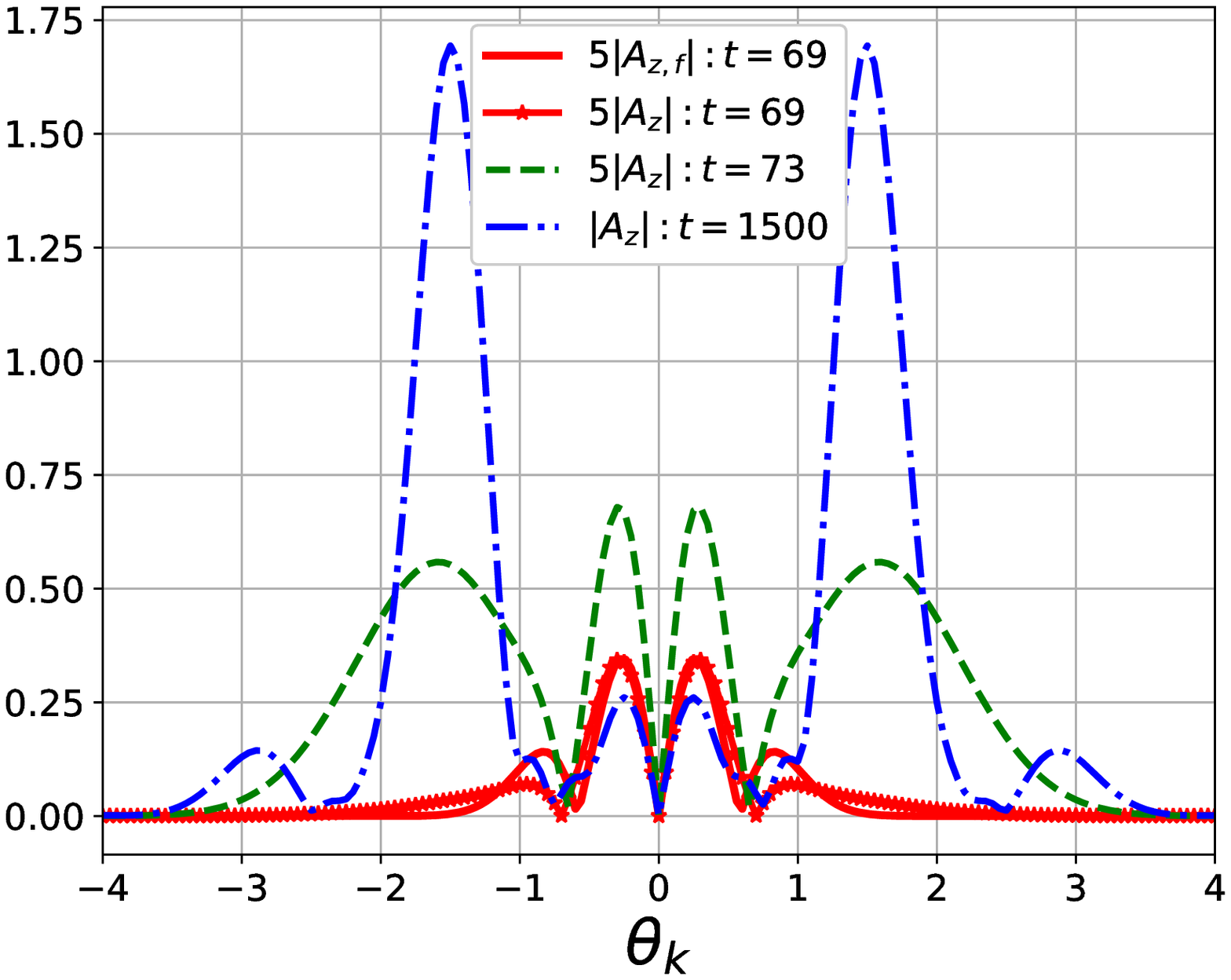}
	\caption{(Color online) Snapshots of the  ETG-ZF  radial spectra at the force-driven ($t=69$), early nonlinear ($t=73$)  and final saturated ($t=1500$) phases. The other parameters are the same as Fig. (\ref{eps:history}).}
\label{eps:snapshots}
\end{figure}

The NLSE model is of integrodifferential nature and generally requires numerical solution.
Figure (\ref{eps:history}) shows the typical time histories for the averaged amplitude $\langle |A_{j}|\rangle=(\int d\theta_{k}|A_{j}|^{2})^{1/2}$ and dimensionless radial wavenumber $\langle|\theta_{j}|\rangle=(\int d\theta_{k} \theta_{k}^{2}|A_{j}|^{2})^{1/2}/\langle |A_{j}|\rangle$ of the ETG and ZF.
One can identify that three stages of the nonlinear evolution of NLSE model. The first stage is early  on before the global ETG mode structure is formed. Although the ZF is being force driven, its spectrum is very sensitive to the specific initial conditions for the ETG and thereby unpredictable.
In the second stage, a global ETG linear mode structure has already been formed, but the ETG nonlinearity is still negligible. In this case, the ZF spectrum can be analytically evaluated as
\begin{eqnarray}
\label{eq:force_zf}
	A_{z,f}=-\frac{i\pi \theta_{k} e^{2\gamma_{g} t}|A_{n,0}|^{2} \textrm{Im}[(\sigma_{r}+\sigma^{2}k_{z}^{2}\rho_{e}^{2})a_{n}]}{4\chi_{z} \sigma_{r}^{5/2}e^{\frac{|\sigma|^{2}}{2\sigma_{r}} k_{z}^{2}\rho_{e}^{2}}[\gamma_{z}(1+d_{z}k_{z}^{2}\rho_{e}^{2})+2\gamma_{g}]},\nonumber
\end{eqnarray}
where $\sigma^{2}=-b_{n}/4c_{n}$, $\delta \Omega=i\gamma_{n}-2c_{n}\sigma$ and $\gamma_{g}=\textrm{Im}(\delta\Omega)$ is the growth rate of the global linear ETG with $A_{n,f}=A_{n,0}\textrm{exp}(-\sigma k_{z}^{2}\rho_{e}^{2}-i\delta\Omega t)$.
The defining feature of the force-driven process, i.e., an $\textrm{exp}(2\gamma_{g}t)$ factor, is readily recognized.
It is worthwhile mentioning that, although the spectral shape is deterministic, the ZF intensity will depend on initial conditions.
Figure (\ref{eps:snapshots}) shows that the predicted ETG-ZF spectral shapes at the force-driven stage ($t=69$) are in qualitative agreement with numerical results. 
%The importance of the forced generation of ZF is, therefore, still sensitive to the details of the adopted initial condition.
%\end{linenumbers}

\begin{figure}[!htp]
\vspace{-0.3cm}
\setlength{\belowcaptionskip}{-0.1cm}
\centering
\includegraphics[scale=0.35]{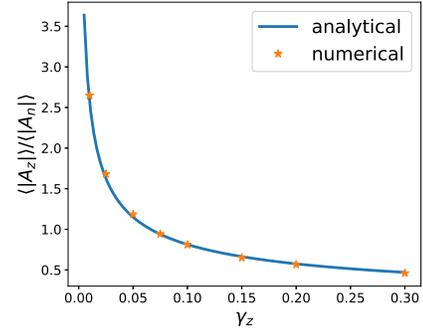}
	\caption{(Color online) The ZF-ETG ratio averaged over time and $\theta_{k}$  versus $\gamma_{z}$. The other parameters are the same as Fig. (\ref{eps:history}).}
\label{eps:AzAn}
\end{figure}

%\begin{linenumbers}
When the ETG grows to the threshold intensity, the spontaneous ZF generation  starts and the system evolves to the nonlinear saturation stage.
As shown in Fig. (\ref{eps:history}), ZFs are initially excited around $|\theta_{k}|\gtrsim |\theta_{z,m}|$ in the early nonlinear state ($t=73$). 
Subsequently, a steady state is gradually reached as ZF spectrum shifts toward $\pm\theta_{z,m}$ and, meanwhile, ETGs are scattered into the linearly stable regime. 
%These features are consisent with the analytical predictions from the four-wave model.
%which is vividly illustrated by the increase in $\langle |\theta_{n}|\rangle$ in the lower panel of Fig. (\ref{eps:history}).
%In the final nonlinear saturation stage,
The effect of long-time-scale ETG-ZF interplay, therefore, is to broaden the ETG radial spectrum, but narrow the ZF spectrum, as vividly illustrated by the snapshots in Fig. (\ref{eps:snapshots}).  
The final steady state is characterized by narrow-band ZFs, and the four-wave model is expected to offer a relevant tool for interpreting numerical results of the complicated NLSE model.
Figure  (\ref{eps:AzAn}) shows that the time-averaged ZF-ETG ratio, computed from Eqs. (\ref{eq:fp_Az}-\ref{eq:fp_A0}) with $\langle A_{z}^{2}\rangle=A_{z,p}^{2}(\theta_{z}=\theta_{z,m})$ and $\langle A_{n}^{2}\rangle=(1+|\gamma_{n}/\gamma_{s}|)A_{0,p}^{2}(\theta_{z}=\theta_{z,m})$, indeed agrees quantitatively with numerical results. 
By taking $W\simeq 0.8$ by inspection of Fig. (\ref{eps:snapshots}), the ZF saturation level $\langle |A_{z}|\rangle=2$ is also in good agreement with the analytical value $|A_{z,p}|\simeq 2.1$.
%Furthermore, noting that the saturation level of ZF is independent of $\gamma_{z}$, the system will be dominated by ZFs in the low collisionality regime, as expected.
Furthermore, in order to give a qualitative estimate for the corresponding electron heat transport level, one can evaluate the quasilinear electron energy flux \cite{lee} approximately as 
$Q_{e}/Q_{gB}\sim \mathcal{O}(|A_{n}|^{2}|k_{\theta}\rho_{e}|)$, where $Q_{gB}=nT_{e}v_{te}\rho_{*}^{2}$ is the gyro-Bohm heat flux.
Therefore, for typical plasma conditions, this suggests that the  electron heat transport caused by final saturated intermediate-scale toroidal ETG turbulence will be $Q_{e}/Q_{gB}\sim \mathcal{O}(0.1)-\mathcal{O}(1)$, with a linear dependence on the collisionality \cite{diamond98}.

%While the exact solutions cannot be obtained explicitly, several actual quantitative features of the ETG-ZF system are readily apparent.

%\section{Conclusions}
%\label{sec:may:01:16:04}
To summarize, we have derived a nonlinear Schr\"{o}dinger equation model for the zonal flow generation in ETG turbulence, by allowing  plasma non-uniformities and properly taking into account the crucial kinetic effects.
It is demonstrated that ZF is easily excited by the  intermediate-scale toroidal ETG turbulence, and the corresponding threshold condition is lower than previous fluid predictions by at least $\mathcal{O}(k_{\theta}^{2}\rho_{e}^{2})$, due to the Navier-Stokes type nonlinearity in ETG dynamics. 
The three-stage evolution of the coherent ETG-ZF system has been addressed.
The parallel decoupling effect is found to be essential for determining the narrow-band ZF in the final saturated state.
Conversely, the ETG spectrum is broadband since the saturation is achieved via scatterings to the high-$\theta_{k}$ stable regime.
Considering typical tokamak parameters, the electron heat transport level expected for the ETG-ZF system is in the range $Q_{e}\lesssim Q_{gB}$ and proportional to the collisionality.
Therefore, ZF generation is an efficient nonlinear mechanism for the isotropization and saturation of the intermediate-scale toroidal ETG turbulence.
These theoretical findings provide an explanation for the recent gyrokinetic simulation results \cite{howard16,colyer,holland17} that show ZF can be important for ETG-driven transport. 

Finally, we note that, the present work is readily extended to include the nonlinear toroidal mode coupling of unstable ETGs.
For short-wavelength ETG turbulence, the fluid description is applicable, previous studies \cite{chen05,lin05} have shown that the toroidal inverse cascade will dominate over the spontaneous ZF excitation, and the resulting turbulence is characterized by streamers.
As energy is transferred  to intermediate-scale turbulence, however, the ETG-ZF interactions become important and ETG turbulence will be isotropized.

%Finally, we remark that the present work can be easily extended to the kinetic region.

%\section{Acknowledgments}
%\label{sec:may:01:16:05}

This work is supported by the Exascale Computing Project (17-SC-20-SC), a collaborative effort of the U.S. Department of Energy Office of Science and the National Nuclear Security Administration.
We thank Prof. Liu Chen, Dr. Yang Chen and Dr. Junyi Cheng for useful conversations.

%\end{linenumbers}

\end{document}